%% file: bg-complenet.tex
\documentclass[runningheads]{llncs}
\usepackage{graphicx}
\usepackage[hyphens]{url}  
\RequirePackage[colorlinks,citecolor=blue,urlcolor=blue]{hyperref}
\usepackage{enumitem}
\usepackage{amsmath}
\usepackage{xcolor}
\usepackage{booktabs}
\usepackage{algorithm}
\usepackage[noend]{algpseudocode}
\usepackage[para]{footmisc} 
\usepackage{sidecap}
\sidecaptionvpos{t}{figure}
\usepackage{bbm}
\makeatletter
\def\BState{\State\hskip-\ALG@thistlm}
\makeatother
\makeatletter
\newcommand{\printfnsymbol}[1]{%
  \textsuperscript{\@fnsymbol{#1}}%
}
\makeatother

\usepackage{caption} 
\begin{document}
\title{An Interpretable Graph-based Mapping of Trustworthy Machine Learning Research}
\titlerunning{Graph-based Mapping of Trustworthy ML Research}
%

%
\author{Noemi Derzsy\thanks{Alphabetical authors.}
\and Subhabrata Majumdar\printfnsymbol{1}
\and Rajat Malik\printfnsymbol{1}
}
\authorrunning{Derzsy et al}
\institute{Data Science and AI Research, AT\&T Chief Data Office
\email{\{nderzsy,subho,rmalik\}@att.com}}
\maketitle              
\begin{abstract}
There is an increasing interest in ensuring machine learning (ML) frameworks behave in a socially responsible manner and are deemed trustworthy. Although considerable progress has been made in the field of Trustworthy ML (TwML) in the recent past, much of the current characterization of this progress is qualitative. Consequently, decisions about how to address issues of trustworthiness and future research goals are often left to the interested researcher. In this paper, we present the first quantitative approach to characterize the comprehension of TwML research. We build a co-occurrence network of words using a web-scraped corpus of more than 7,000 peer-reviewed recent ML papers---consisting of papers both related and unrelated to TwML. We use community detection to obtain semantic clusters of words in this network that can infer relative positions of TwML topics. We propose an innovative fingerprinting algorithm to obtain probabilistic similarity scores for individual words, then combine them to give a paper-level relevance score. The outcomes of our analysis inform a number of interesting insights on advancing the field of TwML research.

\end{abstract}

\input{1-introduction}
\input{2-methods}

\input{3-results}

\input{4-disc}
\input{5-conc}


%
\bibliographystyle{splncs04}
\bibliography{bg-complenet} 

\end{document}

%% file: 1-introduction.tex
\section{Introduction}
With the unprecedented increase in the deployment of machine learning (ML) systems in the real world, there is an increasing need to ensure that such systems behave in a socially responsible manner. Responding to this challenge, in the recent past there has been a plethora of interest from ML researchers and practitioners to develop algorithms and models that embody qualities such as fairness, explainability, privacy, and robustness. This sub-field of ML is often referred by umbrella terms such as Responsible ML or Trustworthy ML \cite{cheng2021socially,trustmltech,xiong2021robust}.

Scientific literature on trustworthy ML (TwML) has grown rapidly in the past few years. While this presents tremendous opportunities for future technical work, there is a lack of codification and characterization of this knowledge base. Considering the interdisciplinary nature of the area and its major venues of publication (e.g. FAccT\footnote{\url{https://facctconference.org}} and AIES\footnote{\url{https://www.aies-conference.com}}), with participation from fields like social sciences and public policy, such mapping is important to not only summarize existing work but also to inform new application areas on their relevance to TwML. To this end, there are a number of quality review and summary articles \cite{cheng2021socially,ChRoth20,MehrabiEtal19}, as well as books \cite{kearns2019ethical}. However, by nature these are qualitative and static, and leave the judgement of a topic of interest being relevant to TwML to the reader.

In this paper, we take a quantitative approach to characterize the comprehension of trustworthy ML research and its position with respect to contemporary scholarly work in the broader field of ML. Our text analytics approach is based on a weighted co-occurrence network of words which occur in the text of more than 7,000 peer-reviewed ML papers published in the last 5 years---both related and unrelated to TwML. We use this network for two purposes. First, we use community detection to obtain semantic clusters of words and infer the relative position of TwML topics. Second, we propose a novel relevance score to quantify the `closeness' of individual words to TwML concepts, then combine these word-level scores in an interpretable manner to obtain paper-level relevance scores.

\paragraph{Related Work.}
Easy availability of bibliographic data has enabled a number of recent studies that aim to map scientific research spaces based on past scholarly work. Such studies span both general science \cite{Fortunato,Zeng}, and specific domains like Physics \cite{Chinazzi,Palmucci}, Bioinformatics \cite{LiBaiYang}, as well as rapidly emerging interest areas like COVID-19 \cite{yeganova2020navigating}. The use of complex network models is extremely popular in these studies. In previous work, network models have been built on data from citations \cite{Portenoy}, author information and collaborations \cite{Chinazzi,CIMINI}, pre-categorized research topics, such as Physics and Astronomy Classification Scheme (PACS) codes \cite{Chinazzi,Palmucci}, or words in the text of papers \cite{LiBaiYang}.

Popular methods for constructing scientific knowledge fall in two broad categories: embedding-based methods, and co-occurrence networks/knowledge graphs. Embedding-based methods, such as \cite{Chinazzi}, typically map entities like words, sentences and documents to a high-dimensional numerical space (using techniques such as \texttt{Word2vec}), then form edges between two entities (such as words, authors, citations) based on their similarity as measured by some similarity metric. On the other hand, the second category of methods build a network directly based on the connectivity patterns of entities. Edges or edge weights in the graph may correspond to specific relationships between entities for knowledge graphs \cite{buscaldi2019mining}, or measures such as co-occurrence indicators/weights \cite{LiBaiYang,Radhakrishnan}.

To the best of our knowledge, the literature lacks a study which maps the research landscape and characterizes existing knowledge of TwML. In this paper we aim to fill this gap.

\paragraph{Goals and contributions.}
Contrasting with the mostly exploratory and inferential nature of studies on other fields of research, we aim for both inference and prediction. Specifically, our goal is to answer the following research questions in a data-driven manner:

\noindent{\bf Q1:} Within the co-occurrence network of words in recent ML literature, can we characterize the relative position of words and concepts as they relate to TwML?

\noindent{\bf Q2:} Can we predict which scientific papers that are {\it not} on TwML topics may be relevant to this sub-field?

\noindent{\bf Q3:} For words or terms not directly pertaining to TwML, can we infer their context similarity with words/terms that do?

To address these questions, our contributions are:
\begin{enumerate}[leftmargin=*]
\setlength\itemsep{0em}
    \item We map the space of recent ML research using a word-level network generated from papers that cover general ML, as well as work specifically on TwML.
    \item We propose a probabilistic fingerprinting method that quantifies the relevance of a word/paper to TwML. Word-level relevance scores are aggregated in an interpretable manner to obtain the paper-level scores.
    \item Taking a broader point of view, we explore the network of words to identify a words that, even though not overtly related to TwML (i.e. unlike terms such as fairness, transparency), are {\it conceptually} related to TwML.
\end{enumerate}
As the nascent area of Trustworthy ML matures, through this paper we aim to initiate the quantitative study of this body of literature to channel future work in interesting directions, as well as create connections with new application areas.

%

%% file: 2-methods.tex
\section{Materials and Methods}
\label{sec:matmeth}

\subsection{Data}
\label{subsec:data}
To begin with, we scraped 7107 papers from the Proceedings of Machine Learning Research\footnote{\url{http://proceedings.mlr.press}} (PMLR) website, that have been presented at peer-reviewed conferences and workshops, and contain ML research covering a breadth of topics. This corpus as part of our analysis ensures we have a diverse dictionary of words and their co-occurrences.

Due to the wide scope of the PMLR corpus, papers that specifically focus on TwML get clubbed together with ones that do not. To create a distinctive set of papers that focus on TwML, we used a two-pronged approach. First, we obtained 221 papers from the ACM Digital Library\footnote{\url{https://dl.acm.org}} that were published in past FAccT conferences (2018--2020), and labeled all of them as TwML-focused. Second, we curated a set of 74 words and terms related to TwML, starting from a list of obvious `seed' words (e.g. `bias',`fairness') and manually iterating to include their variants relevant for TwML (e.g. `algorithmic bias' but not `biased coin') present in the FAccT corpus. We then label a paper in the larger PMLR corpus as TwML-focused if it contains at least one occurrence of any of these words, resulting in 263 more such papers. We label all the other PMLR papers as non-TwML-focused.

{\it Pre-processing.} In a scientific paper, raw data, tables, plots, proofs, and references can all contribute to a noisy dictionary. Authors spend considerable time deciding titles and writing abstracts to make them stand out more. Further, abstracts often present a high-level summary of the research problem and methodology \cite{Anaesth19}. Therefore, for our analysis, we restrict our corpus to only include titles, keywords (when present), and abstracts.

We start with standard text pre-processing steps: convert text to lowercase, remove special and numeric characters, tokenize, remove stop words and single character words, and then finally stem words using the Snowball stemmer\footnote{\url{https://snowballstem.org}}
. 
After pre-processing, we use simple random sampling to split the corpus, assigning 90\% of all papers (i.e., PMLR+FAccT) as training set and the remainder as test set. 


\subsection{Methods}
\label{subsec:meth}
\begin{figure}[t]
    \centering
    \includegraphics[width=\linewidth]{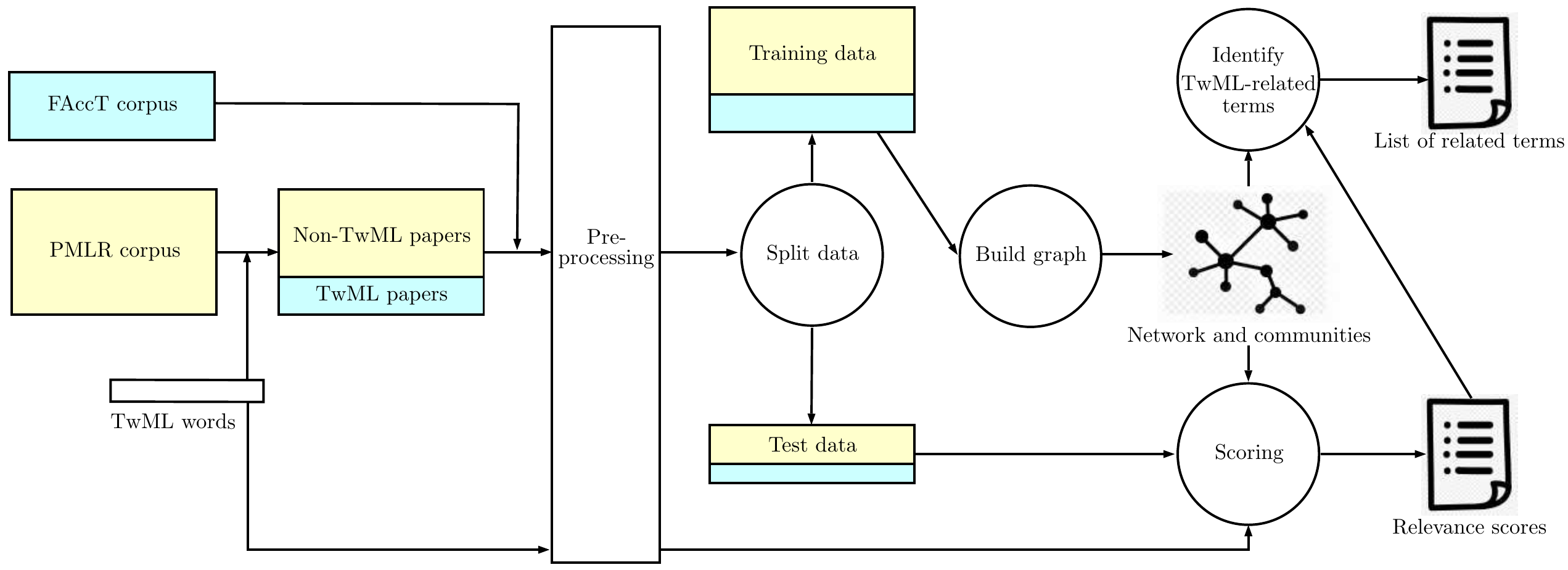}
    \caption{Schematic of the methodology. Blue and yellow indicate papers in a corpus labeled as TwML and non-TwML.}
    \label{fig:schematic}
\end{figure}

Our methodological work has 3 components: (1) building a network of words and detecting communities of similar words, (2) fingerprinting papers as relevant to TwML, and (3) discovery of non-TwML words potentially relevant to the area of TwML. Figure~\ref{fig:schematic} illustrates these processes, and we detail them below.

\paragraph{Network of words.}
Using the pre-processed text from our training corpus, we build a word co-occurrence network by connecting each pair of stemmed words that appear in the same abstract. Connections between words (i.e., nodes) are represented with a weighted edge. The weight reflects co-occurrence---the number of times the pair of words appeared together in an abstract. This construction scheme generates an undirected, weighted network of words. Next, we detect communities in this network and identify which communities our predefined list of TwML words occur in. In order to effectively perform community detection on the network, we use additional cleaning steps to denoise the graph by removing very high-frequency words. We apply a differential edge cutoff: we remove the top 10\% highest connectivity non-TwML words 
and the top 25\% of highest connectivity TwML words that originate from splitting compound words (e.g., `algorithmic bias' $\rightarrow$ `algorithm' and `bias'). Note that this splitting also converts the 74 TwML-specific words into 41 individual stemmed words.
Finally, we use the Louvain community detection algorithm \cite{Blondel_2008} to identify densely connected communities within the above network.

\paragraph{Bi-level fingerprinting.}
We use a novel fingerprinting algorithm (Algorithm~\ref{algo:biasscoring}) to obtain probabilistic similarity scores for individual words or papers.

\begin{algorithm}[h]
\caption{Algorithm for word-level relevance scoring}
\label{algo:biasscoring}
\begin{algorithmic}
\Procedure{scoreword} {word, TwMLwords, Graph}
\If {word {\bf in} TwMLwords}
\State {\bf return} 1;
\Else
\State path = [];
\If {word {\bf in} Graph}
\For {twml {\bf in} TwMLwords}
\State sp = weighted shortest path(word, twml);
\If {sp != \textsc{null}}
\State path.append(sp);
\EndIf
\EndFor
\EndIf
\State {\bf return} path.mean() / path.max()
\EndIf
\EndProcedure
\end{algorithmic}
\end{algorithm}
To begin with, we score each word individually based on its weighted shortest path distance---computed using Dijkstra's algorithm---from TwML words. We calculate the relevance score for a full paper as the weighted average of the word-level scores of all words in that paper, assigning larger weights to words that belong to the same community as TwML words:

\begin{equation}
    s_{\mathrm{paper}} = \frac{\sum_{i=1}^N 
    s_i [w_1 \times \mathbbm{1}_{i \in \mathrm{TC}} +
    w_2 \times \mathbbm{1}_{i \in \mathrm{NTC}}]}
    {\sum_{i=1}^N [w_1 \times \mathbbm{1}_{i \in \mathrm{TC}} +
    w_2 \times \mathbbm{1}_{i \in \mathrm{NTC}}]}.
    \label{eqn:projscore}
\end{equation}
Here, $s_i > 0$ is the relevance score of the $i^\text{th}$ word in the paper. Given weights $w_1 > w_2 \geq 0$, the contribution of a word to the paper-level score is $w_1 s_i$ if it belongs to any of the two communities rich in TwML words (Table~\ref{tab:commtable}; indicated by TwML community, or TC), and $w_2 s_i$ if it belongs to any other community (indicated by non-TwML community, or NTC). To score a paper, we consider the $N$ words in its abstract that yield non-zero scores through Algorithm~\ref{algo:biasscoring}. Finally, the denominator normalizes a paper-level score by the maximum possible value, and the score is set to 0 if all word-level scores are 0 in a paper. If the relevance score of a paper is $\geq 0.5$, then we flag the paper as potentially TwML-related. We use grid search to find optimal values of the weights
: $w_1=3, w_2=0.5$.


Due to the nature of how the above paper-level relevance scores (Eq. \ref{eqn:projscore}) are calculated, our probabilistic fingerprinting method is inherently interpretable. From analyzing the breakdown of a paper-level score into its constituent word-level scores, the user can obtain potential reasonings of why a paper may be (or not) highly relevant to TwML. We discuss this in Sections \ref{sec:res} and \ref{sec:disc}.

\paragraph{Contextual similarity of non-TwML words.}
Our last goal is to expand the existing list of TwML words with additional words that are conceptually related to TwML. The reason for doing this is two-fold. Firstly, in the current work we rely solely on the TwML words as an initial seed list of mostly technical words that are used for multiple purposes. However, expanding this existing list with additional contextually similar words would result in a more inclusive set that can improve the fingerprinting process. Secondly, we wish to identify broad areas of interest for future research using these conceptually similar words.

To this end, we utilize the connectivity information of non-TwML words with TwML words. 
We extract all the direct connections of TwML words, along with their corresponding edge weights, which indicate the strength of their connection. In addition, we score each direct neighbor using Algorithm~\ref{algo:biasscoring}, which informs us on the overall connectivity of that word with TwML words as a whole. Finally, we use upper threshold cutoffs on edge weights and word relevance scores to identify words above the threshold as potentially of interest. 

%% file: 3-results.tex
\section{Results}
\label{sec:res}

\paragraph{Network of words.}
Only about 7\% (484 out of 7328) of all papers are TwML-related. Previous studies have empirically observed that complex methods such as knowledge graphs or high-dimensional numeric embeddings are less reliable for characterizing rare concepts or terms \cite{ManningBook,Tacchella}. Because of this rarity issue of TwML papers, we use a word co-occurrence network in place more sophisticated methods. The resulting network contains 10,698 nodes and 254,347 edges. 

%
\begin{table}[t]
\centering
\scalebox{0.85}{
\begin{tabular}{|p{2cm}|p{2.2cm}|p{9cm}|}
\hline
    {\bf Community size} & {\bf Number of TwML words} & {\bf TwML words} \\\hline
    1127 & 26 & sensit, bias, decis, constraint, impact, group, remov, discrimin, attribut, demograph, fair, gender, implicit, interpret, mitig, pariti, treatment, unfair, criteria, dispar, sex, subgroup, transpar, crimin, racial, justic \\\hline
    405 & 7 & differenti, mechan, privaci, privat, concern, individu, preserv \\\hline
    488 & 2 & metric, definit \\\hline
    301 & 2 & account, procedur \\\hline
    1228 & 1 & discoveri \\\hline
    980 & 1 & trustworthi \\\hline
    250 & 1 & hindsight \\\hline
    748 & 1 & unbias \\\hline
    \end{tabular}}
    \caption{TwML words in communities. The first two rows contain the bulk of TwML-related words. The second row can be interpreted as a community relating to differential privacy.}
    \label{tab:commtable}
\end{table}

\begin{SCfigure}
    \centering
    \caption{Network of ML research space constructed from the PMLR+FAccT corpus. Each node represents a term, and each edge represents the number of times a pair  of  words co-occur in an abstract. We highlight the two communities containing most TwML words---nodes are colored according to community membership. TwML words are highlighted separately per their subject area.}
    \label{fig:graphpic}
    \includegraphics[width=.65\columnwidth]{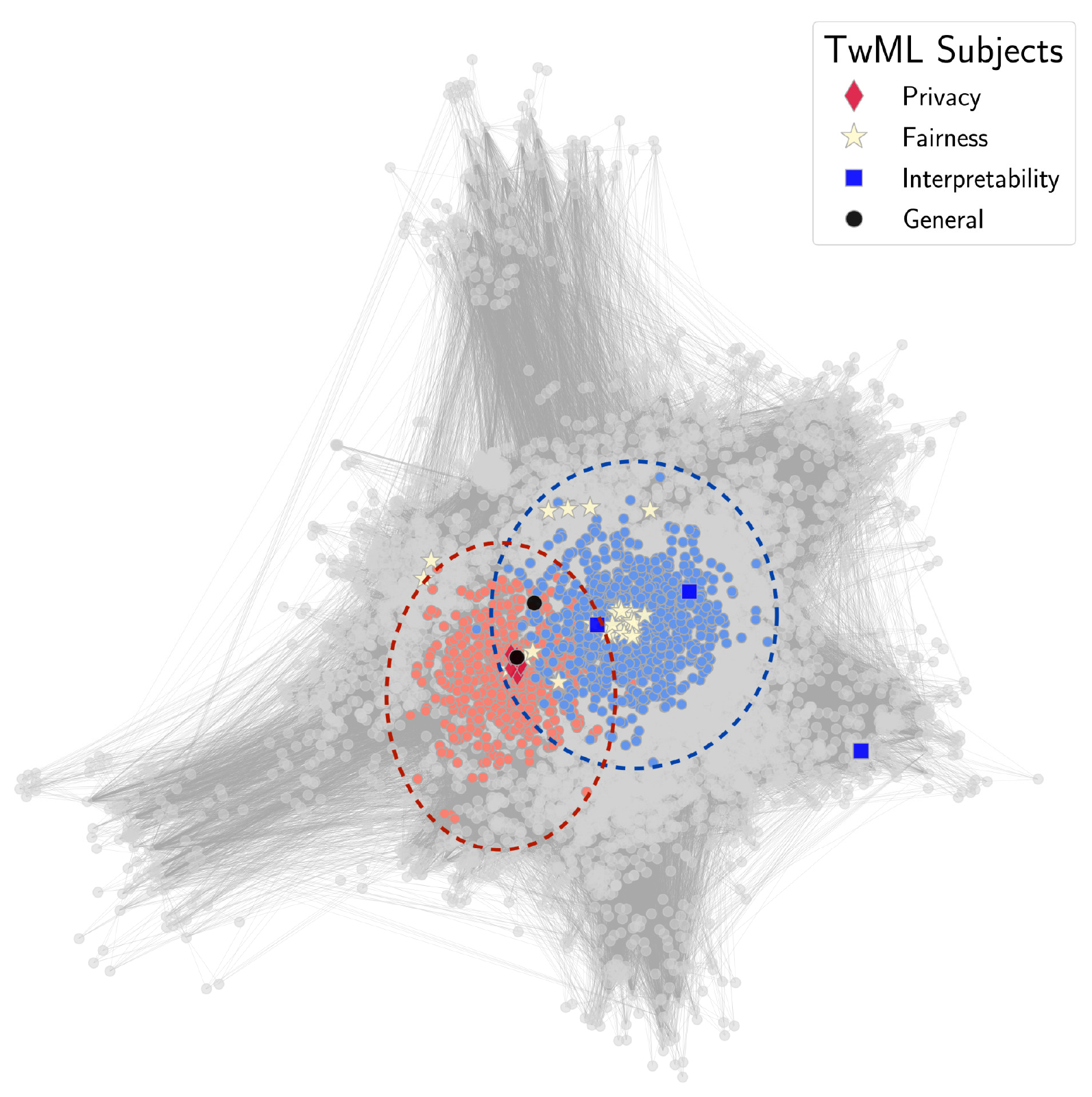}
\end{SCfigure}

The community detection algorithm generated 25 communities, with a modularity score of $0.33$. As given in Table~\ref{tab:commtable}, TwML-related words are concentrated in two communities. Among them, seven words that are mostly related to Differential Privacy (DP) separate from the rest into one community (second row in Table~\ref{tab:commtable}). Another community of 1127 words contains 26 other TwML-specific words. For convenience we shall refer to these communities as DP and non-DP community, respectively. The remaining 8 TwML words---which are mostly ambiguous such as `metric' or `procedur' or general such as `trustworthi'---get distributed across 6 communities.

\begin{SCtable}[][t]
\centering
\scalebox{.9}{
    \begin{tabular}{|l|c|c|c|c|}\hline
{\bf Corpus}   & {\bf AUC} & {\bf Precision} & {\bf Recall} & {\bf F1 score} \\\hline
    PMLR        & 0.81     & 0.42       & 0.81  & 0.55  \\
    FAccT       & --       & 1          & 0.88  & 0.94     \\
    Overall     & 0.82     & 0.47       & 0.82  & 0.6  \\\hline
    \end{tabular}
}
    \caption{Performance of paper-level scoring. AUC = Area Under Curve uses paper-level scores (Eq.~\ref{eqn:projscore}). For other metrics, we use an upper cutoff of 0.5.}
    \label{tab:scoringtable}
 \end{SCtable}
Figure~\ref{fig:graphpic} visualizes the overall network, focusing on the two TwML-specific communities. We categorize the TwML words into four subject-based categories:
\begin{itemize}[leftmargin=*]
\setlength\itemsep{0em}
    \item {\bf Privacy:} `privaci', `differenti', `privat', `guarantee’, `concern’,`preserv',
    \item {\bf Interpretability:} `transpar',`interpret',`account',
    \item {\bf General:} `trustworthi', `mechan',`algorithm’,`data’,
    \item {\bf Fairness:} all others.
\end{itemize}
From the relative position of words in each category in Figure~\ref{fig:graphpic}, it is evident that a number of privacy-specific and fairness-specific words cluster together, and these two clusters are well-separated from each other. 

\paragraph{Fingerprinting of papers.}
Because of the probabilistic nature of our fingerprinting process, it can be used to classify whether or not a paper is related to TwML. Table~\ref{tab:scoringtable} presents the results across different metrics and the two corpus.

Our method exhibits good recall values across the two corpuses. The precision in the PMLR corpus---hence the overall precision, as it forms a large proportion of the overall set of paper---is low. This is an indication that there are probably a number of papers that do not contain our pre-specified TwML words, but may be related to this subject based on their contents. Note that since all papers in the FAccT corpus are labeled as TwML-related, area under curve (AUC) does not exist for this category, and it exhibits a perfect precision.

\begin{figure}[t]
    \centering
    \includegraphics[width=.24\textwidth]{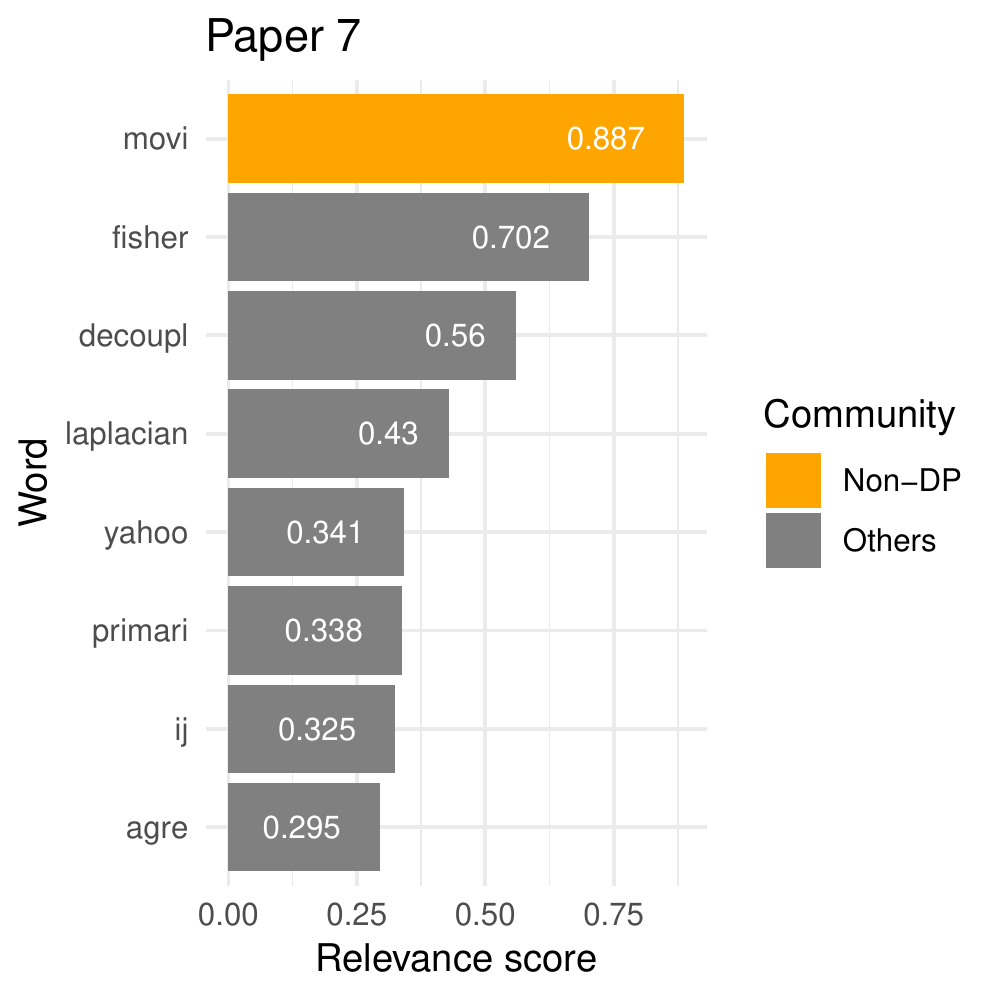}
    \includegraphics[width=.24\textwidth]{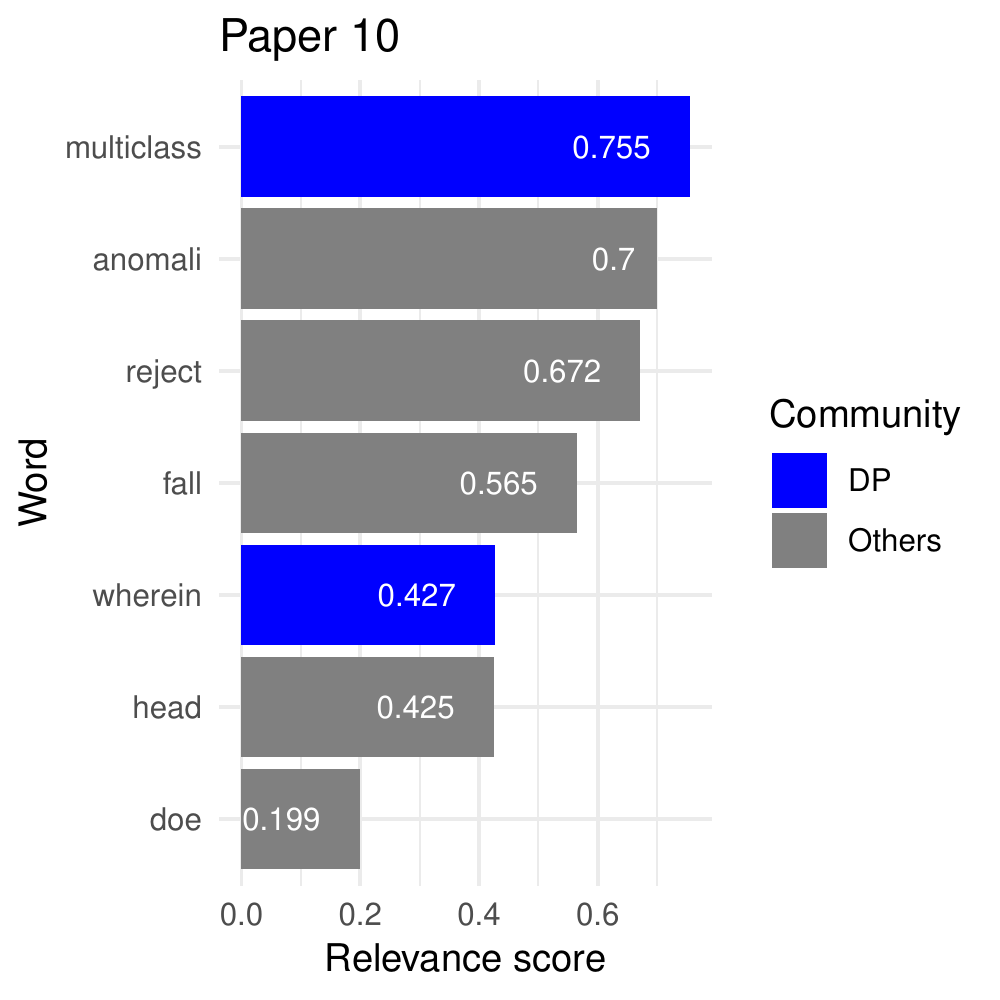}
    \includegraphics[width=.24\textwidth]{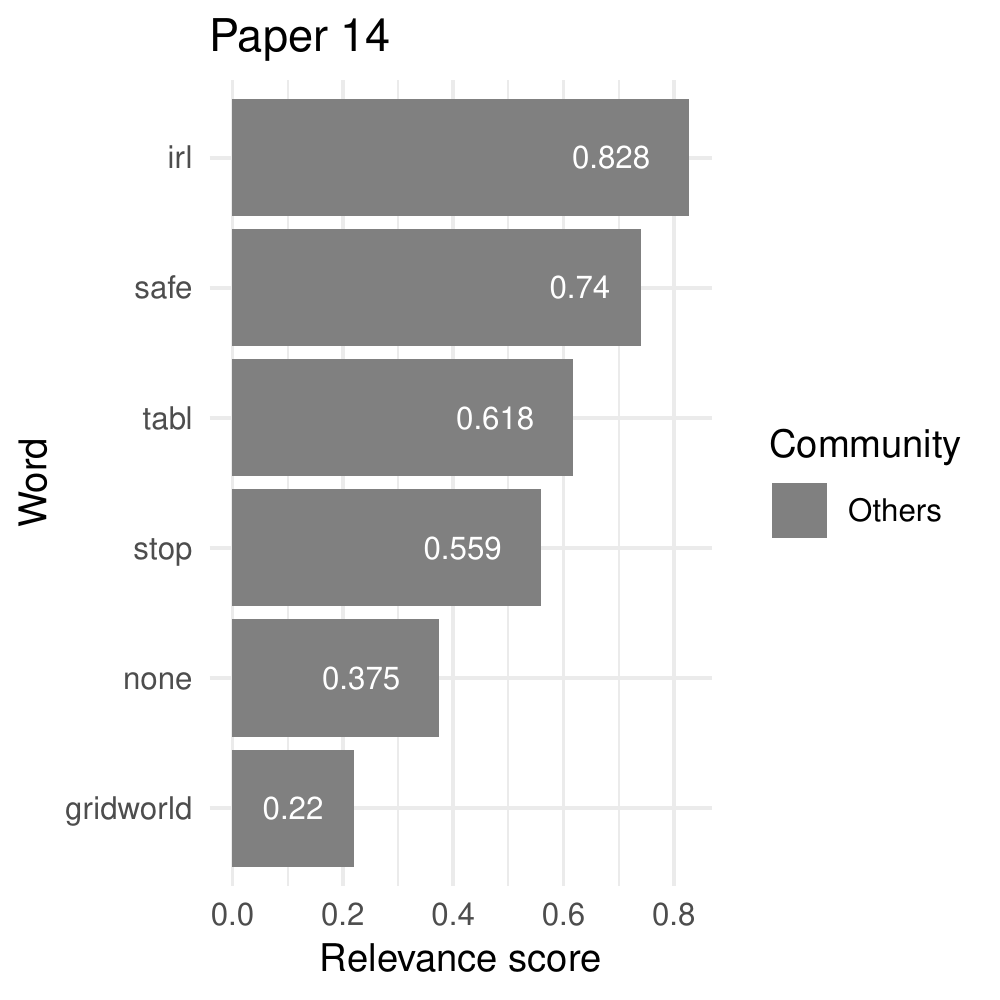}
    \includegraphics[width=.24\textwidth]{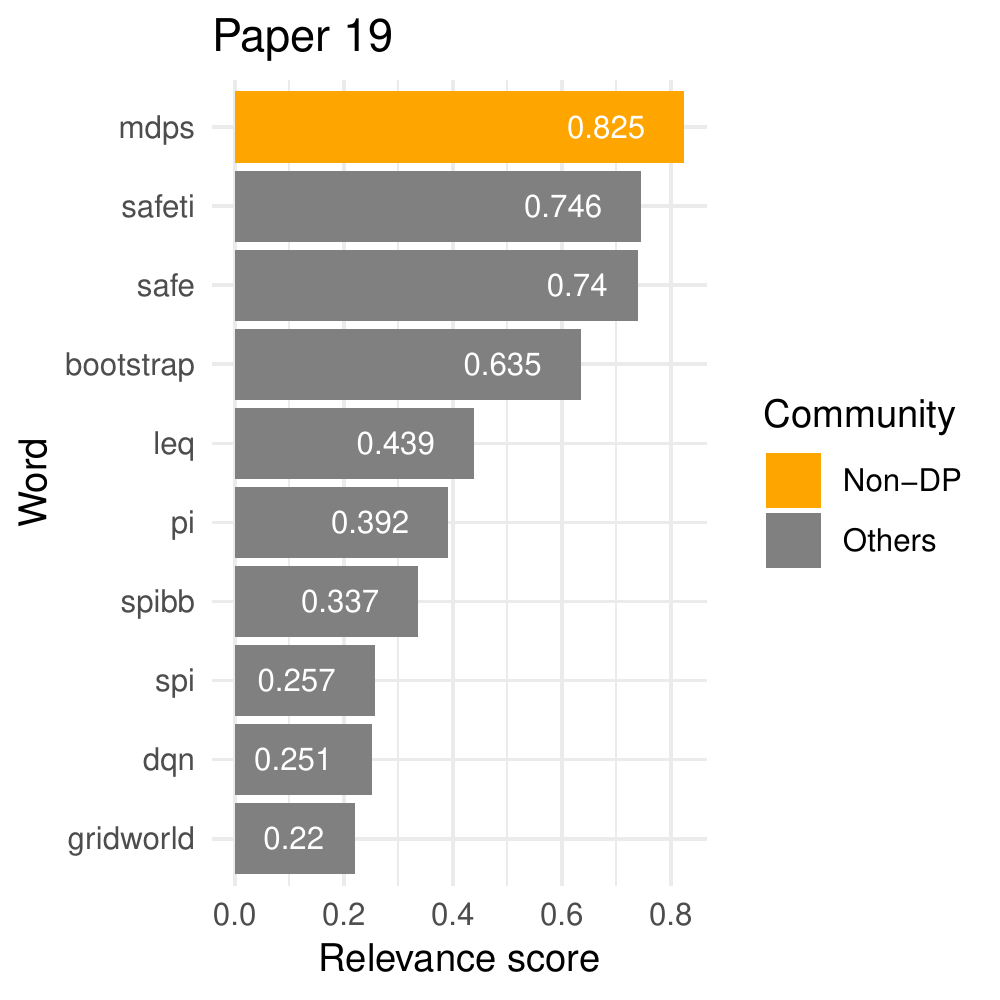}
    \caption{Word level relevance scores for selected papers.}
    \label{fig:wordscores}
\end{figure}

We present non-TwML papers with highest relevance scores in Table~\ref{tab:toppapers}, and the word-level relevance scores for selected papers in Figure~\ref{fig:wordscores}. A number of papers in Table~\ref{tab:toppapers} are on topics that have received less attention in TwML literature \cite{GongReview,MehrabiEtal19}, such as reinforcement learning, active learning, bandit algorithms, and outlier detection. The word-level scores (Figure~\ref{fig:wordscores}) give interpretability to the paper-level fingerprinting. As an example, paper 7 in Table~\ref{tab:toppapers} \cite{Osting13} gets a high score because of the word `movi' from to the non-DP community, and `fisher' which belongs to neither of the two TwML-word rich communities. Contextualizing these words, a fisher information-based approach similar to \cite{Osting13} may be relevant for obtaining fairly calibrated movie ratings and recommendations \cite{Steck}.

\begin{table}[t]
\centering
    \scalebox{.8}{
    \begin{tabular}{c|p{13cm}l}
    \hline
     {\bf Index} & {\bf Paper title}    & {\bf Score} \\\hline
    1 & Sparse Reinforcement Learning via Convex Optimization                                                   & 0.72       \\
    2 &     Boosting with Online Binary Learners for the Multiclass Bandit Problem                                  & 0.71       \\
    3 &     Dirichlet Process Mixtures of Generalized Linear Models                                                 & 0.7       \\
    4 &     Optimal $\delta$-Correct Best-Arm Selection for Heavy-Tailed Distributions                                 & 0.69       \\
    5 &     Lifted Weight Learning of Markov Logic Networks Revisited                                               & 0.7        \\
    6 & Efficient Computation of Updated Lower Expectations for Imprecise Continuous-Time Hidden Markov Chains  & 0.64       \\
    7 &     Enhanced statistical rankings  via  targeted data collection                                            & 0.62       \\
    8 &     Multi-Observation Elicitation                                                                           & 0.6       \\
    9 &     Spotlighting Anomalies using Frequent Patterns                                                          & 0.6       \\
    10 &     Class Proportion Estimation with Application to Multiclass Anomaly Rejection                            & 0.57       \\
    11 &     Exact Subspace Segmentation and Outlier Detection by Low-Rank Representation                            & 0.57       \\
    12 &     Wasserstein Propagation for Semi-Supervised Learning                                                    & 0.56       \\
    13 &     Multitask Principal Component Analysis                                                                  & 0.56       \\
    14 &     Risk-Aware Active Inverse Reinforcement Learning                                                        & 0.56       \\
    15 &     Optimal Densification for Fast and Accurate Minwise Hashing                                             & 0.55       \\
    16 &     A Bayesian Approach for Inferring Local Causal Structure in Gene Regulatory Networks                    & 0.54       \\
    17 &     Lifting high-dimensional non-linear models with Gaussian regressors                                     & 0.52       \\
    18 &     Qualitative Multi-Armed Bandits: A Quantile-Based Approach                                              & 0.51       \\
    19 &     Safe Policy Improvement with Baseline Bootstrapping                                                     & 0.51       \\
    20 &     Cooperative Online Learning: Keeping your Neighbors Updated                                             & 0.51       \\
    21 &     Analysis of Empirical MAP and Empirical Partially Bayes: Can They be Alternatives to Variational Bayes? & 0.5       \\
    22 &     Tree-Based Inference for Dirichlet Process Mixtures & 0.5       \\
    23 &     Sequence Prediction Using Neural Network Classifiers & 0.5       \\
    24 &     Variance Reduction for Faster Non-Convex Optimization & 0.5       \\
    25 &     Stochastic Variance Reduction for Nonconvex Optimization & 0.5       \\\hline
    \end{tabular}}
    \caption{Top 25 papers with highest fingerprinting scores.}
    \label{tab:toppapers}
\end{table}

\paragraph{Contextual similarity.}
To expand the existing list of TwML words with additional conceptually related words, we use the edge weights and relevance scores of words that are direct neighbors of a TwML word in our co-occurrence network to identify the appropriate threshold cutoff. 
Filtering for words that share at least one edge of weight $\ge 100$ with a TwML word, and have a relevance score $\ge 0.5$ resulted in a subset of 290 words. Words in this list can be further assessed for their significance. Table~\ref{tab:commtable1} highlights 10 such words.


%% file: 4-disc.tex
\section{Discussion}
\label{sec:disc}
A number of interesting insights come out from the above analysis. 

\paragraph{Network of words.}
The differential distribution of TwML words within communities, as observed in Table \ref{tab:commtable}, indicates that TwML papers tend to focus more on certain lines of research, methods or applications than others. In the context of ML bias and fairness, this is echoed by the review article of \cite{MehrabiEtal19}. They observed that addressing group fairness in classification problems has received disproportionately high interest compared to other fairness categories (e.g. individual fairness, subgroup fairness) and types of methods (e.g. clustering, graph embedding); see Table 7 therein. Within the TwML words, Differential Privacy (DP)-specific words and those related to fairness and transparency group separately into two different communities. A potential reason for this may be that DP is a comparatively older research area, and has seen more theoretical developments than relatively new topics like fairness or transparency.

\paragraph{Paper-level fingerprinting.}
All papers in Table~\ref{tab:toppapers} with high relevance scores are on comparatively complex algorithms. A number of these areas have been heavily researched of late, such as reinforcement learning (RL; papers 1,14,19), bandit problems (2,4,18), anomaly detection (2,9,10,11), representation learning (11,13,15), multitask problems (2,8,10,13), dirichlet process (3,22), and nonconvex optimization (24,25).

The word-level breakdown of relevance scores (Figure~\ref{fig:wordscores}) gives further insights into how the concepts in these papers may be related to TwML. Top scores for papers 7 and 19 come from TwML-words that belong to the non-DP community. Looking into their subject matters, paper 7 \cite{Osting13} studies statistical ranking for dependent network data. Interestingly, a very recent paper that is not in our analyzed corpus studied the problem of applying fairness constraints on node ranks in a graph \cite{Krasanakis}. Paper 19 is on safe policy improvement in RL \cite{spi}. Safe policies in RL refer to policies that maximize expected return in problems where ensuring certain safety constraints is important alongside satisfactory performance \cite{GarciaSPRL}. In the context of ML fairness, safe policies can potentially be policies that satisfy equitable performance guarantees for sensitive demographic subgroups.

\begin{table}[t]
\centering
\begin{tabular}{|l|l|l|l|l|l|l|l|}
\hline
    {\bf Word} & {\bf Weight} & {\bf Score} & {\bf Community}
&   {\bf Word} & {\bf Weight} & {\bf Score} & {\bf Community} \\\hline
    race & 354 & 0.98 & Non-DP 
&   physiolog & 252 & 0.81 & Others \\
    drug & 324 & 0.88 & Others 
&   censor & 180 & 0.78 & Non-DP \\ 
    tamper & 324 & 0.88 & Non-DP 
&   facial & 198 & 0.76 & Others \\
    stereotyp & 318 & 0.87 & Non-DP 
&   secur & 177 & 0.70 & Non-DP \\ 
    membership & 222 & 0.82 & Non-DP 
&   skin & 180 & 0.67 & Non-DP \\ 
\hline
\end{tabular}
\caption{Contextually similar words to TwML.}
\label{tab:commtable1}
\end{table}
\begin{table}[t]
\centering
    \begin{tabular}{|l|ll|l|ll|}
    \hline
    {\bf Word}       & {\bf Score} & {\bf Community} &
    {\bf Word}       & {\bf Score} & {\bf Community} \\\hline
    movi       & 0.887 & Non-DP    &    vb         & 0.729 & Non-DP    \\
    dp         & 0.825 & DP        &    triplet    & 0.646 & Non-DP    \\
    mdps       & 0.825 & Non-DP    &    chi        & 0.645 & DP        \\
    membership & 0.815 & Non-DP    &    ordinary   & 0.627 & DP        \\
    multiclass & 0.755 & DP        &    opt        & 0.604 & DP        \\\hline
    \end{tabular}
    \caption{Top 10 Words with highest scores in either of the TwML communities.}
    \label{tab:topwordinterp}
\end{table}

In Table~\ref{tab:topwordinterp}, we summarize the words with highest scores among words that occur in any of the 25 papers in Table~\ref{tab:toppapers}, and belong to either the DP or non-DP community. Among words belonging to the DP community, `multiclass' is interesting. After a small number of papers in the early 2010's \cite{Pathak,Sazonova}, multiclass problems in DP have started to receive more attention recently \cite{senekane2019differentially}. Words in the non-DP cluster, on the other hand, refer to methods or algorithms---`mdps' is Markov Decision Processes, `vb' is variational bayes, and `triplet' is triplet loss. Each of these categories are contextual to ML fairness or explainability. For example, \cite{helwegen2020improving,Madras} incorporate causality and fairness notions in ML models using variational inference. Russell and Santos \cite{RussellS19} explains reward functions in MDPs by building a classification model with rewards as outputs. A recent preprint \cite{serna2020sensitiveloss} applies the triplet loss in the context of fairness.

\paragraph{Contextual similarity.}
A large number of `similar' words that are heavily connected with TwML words neither (a) pertain to algorithms or methods, nor (b) belong to the DP community. Table~\ref{tab:commtable1} presents ten such words. In contrast to words highly important to fingerprinting of papers (Table~\ref{tab:topwordinterp}), these similar words mostly refer to application aspects of fairness (`race', `stereotyp', `facial', `skin'), privacy and security (`tamper',`membership',`secur'), as well as other practical issues (`drug',`physiolog',`censor'). This potentially suggest two things. Firstly, application-oriented keywords are closely associated with TwML terms, and should be used to characterize the research landscape of this interdisciplinary field. Secondly, such application areas may foster new connections with TwML topics, especially the ones each such word relates to.

%% file: 5-conc.tex
\section{Conclusion}
\label{sec:conc}
In this paper, we present the first quantitative study of the trustworthy ML research space. Using network analysis methods we identify the similarity and clustering patterns of TwML vs.~non-TwML words, propose a novel fingerprinting method to predict which papers may be related to TwML, and provide word-level contextual similarity insights. As indicated by Table \ref{tab:toppapers}, Figure~\ref{fig:wordscores}, and Table \ref{tab:topwordinterp}, potential areas of future exploration include multiclass problems in differential privacy, and work that focus on fairness and transparency aspects of newer research areas in broader ML. Contextually similar non-TwML words in Table \ref{tab:commtable1} suggest the need for more practice-oriented work in this field, which recent studies have acknowledged \cite{cheng2021socially,sixsteps}. 

Through this work, we hope to motivate further quantitative characterization of TwML literature. As examples, a higher proportion of content instead of only title, abstract, and keywords may be used. The document corpus being analyzed can be specifically tailored to the end goals of the analysis (e.g. inference vs. prediction, explore new connections between theoretical vs. applied topics). Such explorations will facilitate and guide future ML research by identifying methodological gaps, as well as create novel opportunities for applying existing analytical techniques in new practical problems.

%% file: bg-complenet.bbl
\begin{thebibliography}{10}
\providecommand{\url}[1]{\texttt{#1}}
\providecommand{\urlprefix}{URL }
\providecommand{\doi}[1]{https://doi.org/#1}

\bibitem{Blondel_2008}
Blondel, V.D., et~al.: {Fast unfolding of communities in large networks}. J.
  Stat. Mech. Theory and Experiment  \textbf{2008}(10),  P10008 (2008)

\bibitem{buscaldi2019mining}
Buscaldi, D., et~al.: Mining scholarly data for fine-grained knowledge graph
  construction. In: CEUR Workshop Proceedings. vol.~2377, pp. 21--30 (2019)

\bibitem{cheng2021socially}
Cheng, L., et~al.: Socially responsible ai algorithms: Issues, purposes, and
  challenges. arXiv:2101.02032  (2021)

\bibitem{Chinazzi}
Chinazzi, M., et~al.: {Mapping the physics research space: a machine learning
  approach}. EPJ Data Sci.  \textbf{8}(33) (2019)

\bibitem{ChRoth20}
Chouldechova, A., Roth, A.: {A Snapshot of the Frontiers of Fairness in Machine
  Learning}. Commun. ACM  \textbf{63},  82--89 (2020)

\bibitem{CIMINI}
Cimini, G., Zaccaria, A., Gabrielli, A.: {Investigating the interplay between
  fundamentals of national research systems: Performance, investments and
  international collaborations}. J. Informetrics  \textbf{10}(1),  200 -- 211
  (2016)

\bibitem{Fortunato}
Fortunato, S., et~al.: Science of science. Science  \textbf{359}(6379),
  eaao0185 (2018)

\bibitem{GarciaSPRL}
Garc{{\'i}}a, J., Fern{{\'a}}ndez, F.: {A Comprehensive Survey on Safe
  Reinforcement Learning}. J. Mach. Learn. Res.  \textbf{16}(42),  1437--1480
  (2015)

\bibitem{GongReview}
{Gong}, M., et~al.: {A Survey on Differentially Private Machine Learning
  [Review Article]}. IEEE Comput. Intell. Mag.  \textbf{15}(2),  49--64 (2020)

\bibitem{helwegen2020improving}
Helwegen, R., et~al.: Improving fair predictions using variational inference in
  causal models. arXiv:2008.10880  (2020)

\bibitem{kearns2019ethical}
Kearns, M., Roth, A.: {The Ethical Algorithm: The Science of Socially Aware
  Algorithm Design}. Oxford University Press, Incorporated (2019)

\bibitem{Krasanakis}
Krasanakis, E., et~al.: {Applying Fairness Constraints on Graph Node Ranks
  Under Personalization Bias}. In: Complex Networks {\&} Their Applications IX.
  pp. 610--622. Springer International Publishing (2021)

\bibitem{LiBaiYang}
Li, T., et~al.: {Co-Occurrence Network of High-Frequency Words in the
  Bioinformatics Literature: Structural Characteristics and Evolution}. Appl.
  Sci.  \textbf{8}(10), ~1994 (2018)

\bibitem{Madras}
Madras, D., et~al.: Fairness through causal awareness: Learning causal
  latent-variable models for biased data. In: FAT{\*}-2019. p. 349–358 (2019)

\bibitem{ManningBook}
Manning, C.D., {Sch\"utze}, H.: {Foundations of Statistical Natural Language
  Processing}. MIT Press, first edn. (1999)

\bibitem{MehrabiEtal19}
Mehrabi, N., et~al.: {A Survey on Bias and Fairness in Machine Learning}.
  arXiv:1908.09635  (2019)

\bibitem{sixsteps}
Mills, S., et~al.: {Six Steps to Bridge the Responsible AI Gap} (2020),
  \url{https://www.bcg.com/publications/2020/six-steps-for-socially-responsible-artificial-intelligence}

\bibitem{Osting13}
Osting, B., et~al.: Enhanced statistical rankings via targeted data collection.
  ICML-2013 pp. 489--497 (2013)

\bibitem{Palmucci}
Palmucci, A., et~al.: {Where is your field going? A machine learning approach
  to study the relative motion of the domains of physics}. PLoS ONE
  \textbf{15}(6),  e0233997 (2020)

\bibitem{Pathak}
Pathak, M.A., Raj, B.: {Large Margin Multiclass Gaussian Classification with
  Differential Privacy}. In: PSDML-2010. p. 99–112 (2010)

\bibitem{Portenoy}
Portenoy, J., et~al.: Leveraging citation networks to visualize scholarly
  influence over time. Front. Res. Metr. Anal  \textbf{2}, ~8 (2017)

\bibitem{Radhakrishnan}
Radhakrishnan, S., et~al.: Novel keyword co-occurrence network-based methods to
  foster systematic reviews of scientific literature. PLoS ONE  \textbf{12}(9),
   e0185771 (2017)

\bibitem{RussellS19}
Russell, J., Santos, E.: Explaining reward functions in markov decision
  processes. In: Proceedings of the Thirty-Second International Florida
  Artificial Intelligence Research Society Conference, Sarasota, Florida, USA,
  May 19-22 2019. pp. 56--61 (2019)

\bibitem{Sazonova}
Sazonova, V., Matwin, S.: {Combining Binary Classifiers for a Multiclass
  Problem with Differential Privacy}. Trans. Data Privacy pp. 51--70 (2014)

\bibitem{senekane2019differentially}
Senekane, M.: Differentially private image classification using support vector
  machine and differential privacy. Mach. Learn. Knowl. Extr.  \textbf{1}(1),
  483--491 (2019)

\bibitem{serna2020sensitiveloss}
Serna, I., et~al.: {SensitiveLoss: Improving Accuracy and Fairness of Face
  Representations with Discrimination-Aware Deep Learning}. arXiv:2004.11246
  (2020)

\bibitem{spi}
Sim\~ao, T.D., Spaan, M.: {Safe Policy Improvement with Baseline Bootstrapping
  in Factored Environments}. AAAI-2019 pp. 4967--4974 (2019)

\bibitem{Steck}
Steck, H.: Calibrated recommendations. In: RecSys-2018. p. 154–162 (2018)

\bibitem{Tacchella}
Tacchella, A., et~al.: {Novel keyword co-occurrence network-based methods to
  foster systematic reviews of scientific literature}. PLoS ONE
  \textbf{12}(9),  e0185771 (2017)

\bibitem{trustmltech}
Toreini, E., et~al.: {The Relationship between Trust in AI and Trustworthy
  Machine Learning Technologies}. In: FAT{\*}-2020. p. 272–283 (2020)

\bibitem{Anaesth19}
Tullu, M.S.: {Writing the title and abstract for a research paper: Being
  concise, precise, and meticulous is the key}. Saudi J. Anaesth.
  \textbf{13}(Suppl 1),  S12--S17 (2019)

\bibitem{xiong2021robust}
Xiong, P., et~al.: {Towards a Robust and Trustworthy Machine Learning System
  Development}. arXiv:2101.03042  (2021)

\bibitem{yeganova2020navigating}
Yeganova, L., et~al.: {Navigating the landscape of COVID-19 research through
  literature analysis: A bird's eye view}. arXiv:2008.03397  (2020)

\bibitem{Zeng}
Zeng, A., et~al.: {The science of science: From the perspective of complex
  systems}. Phys. Rep.  \textbf{714-715},  1 -- 73 (2017)

\end{thebibliography}
